\documentclass[aip,reprint]{revtex4-1}

\usepackage{amsmath,amssymb}
\usepackage{graphicx}
\usepackage{dcolumn}
\usepackage{bm}

\usepackage[utf8]{inputenc}
\usepackage[T1]{fontenc}
\usepackage{mathptmx}
\usepackage{etoolbox}
\usepackage{color}

\makeatletter
\def\@email#1#2{%
 \endgroup
 \patchcmd{\titleblock@produce}
  {\frontmatter@RRAPformat}
  {\frontmatter@RRAPformat{\produce@RRAP{*#1\href{mailto:#2}{#2}}}\frontmatter@RRAPformat}
  {}{}
}%
\makeatother
\begin{document}


\title{Directional drying of a colloidal dispersion: quantitative description with water potential measurements using water clusters in a   poly(dimethylsiloxane) microfluidic  chip}
\author{Hrishikesh Pingulkar}
\author{Sonia Maréchal}%
\author{Jean-Baptiste Salmon}
\affiliation{CNRS, Solvay, LOF, UMR 5258, Université de Bordeaux,178 av. Schweitzer, Pessac, 33600}%

\date{\today}

\begin{abstract}
We  have developed  a poly(dimethylsiloxane) (PDMS) microfluidic chip to study the directional drying of a colloidal dispersion confined in a channel. Our measurements on a dispersion of silica nanoparticles  
once again revealed the phenomenology commonly observed for such systems: the formation of a porous solid  with linear  growth in the channel at short times, slowing down at longer times as the evaporation rate decreases. The growth of the solid is also accompanied by mechanical stresses that are released by the delamination of the solid  from the channel walls and the formation of cracks.  In addition to these observations, we  report original measurements using hydrophilic filler  in the PDMS formulation used (Sylgard-184). When the PDMS matrix is in contact with water, water molecules pool around these hydrophilic sites,  resulting in the formation of  microscopic water clusters whose size depends on the  water potential $\psi$.
In our work, we have used these water clusters to  estimate the   water potential  profile in the channel as the porous solid grows.   Using a transport model that also takes into account solid delamination in the channel, we then linked these water potential measurements to the hydraulic permeability of the porous solid. These measurements finally enabled us to show that the slowdown in the evaporation rate is due to the invasion of the porous solid by air/water nanomenisci at a critical capillary pressure $\psi_\text{cap}$. 
\end{abstract}

\maketitle

\section{Introduction}
\label{sec:sample1}
\begin{figure}[htbp]
    \centering
    \includegraphics{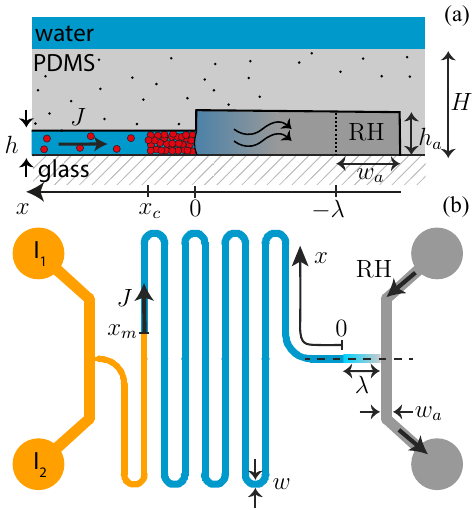}
    \caption{(a) 
 Sketch of confined directional drying   of a colloidal dispersion, and 
 cross-sectional view along the dashed line of the microfluidic  chip shown in (b).  Dimensions: $h=30$, $h_a = 70$, $w_a=300$, $\lambda=500$, and $H\simeq 200~\mu$m.  The water bath on top of the PDMS layer prevents pervaporation. Colloids are shown in red, with black dots representing the water clusters in PDMS, see Section~\ref{sec:cluster}.
 (b) Schematic view  of the microfluidic  chip.
 Blue 
indicates water, yellow fluorinated oil. I$_1$ and I$_2$ are the two liquid inlets. Air flows at a controlled relative humidity $\mathrm{RH}$ in the gray channel (width $w_a$). Evaporation induces a flow at a rate 
$J = -\dot{x}_m$ in the main channel ($w=100~\mu$m)  as long as the air/water meniscus remains pinned at $x=0$.}\label{fig:SetupAirFlow}
\end{figure}

The drying of  colloidal dispersions, a common step in many manufacturing processes, remains a fascinating subject of study whose understanding involves many aspects from the physico-chemistry of colloids, transport phenomena, to the mechanics of porous solids~\cite{Russel2011,Routh2013,GoehringL2015,Bacchin2018}.
For a general understanding of these phenomena, many groups have focused on model cases, such as the confined directional drying~\cite{Allain:95,Dufresne:03,Dufresne:06}. In such experiments, a dilute dispersion is confined within a capillary or Hele-Shaw cell, with cross-sectional dimensions  $<100~\mu$m, see Fig.~\ref{fig:SetupAirFlow}a for a  schematic view. When the air/dispersion meniscus is pinned at the outlet of the cell, solvent evaporation induces a flow at a rate $J$ (m/s) due to mass conservation. This flow then accumulates the colloids at the end of the cell until a consolidated  solid is formed. For rigid colloids, the porosity of the solid does not stop solvent evaporation, even for nanoparticles as small as a few nm~\cite{Dufresne:03,Dufresne:06}, and the solid invades the cell at a rate $\dot{x}_c$. At longer times, the evaporation-induced flow through the porous solid leads to mechanical stresses that are released by instabilities: formation of shear bands~\cite{Boulogne:14,Kiatkirakajorn:15,Yang2015},  delamination of the solid from the cell walls~\cite{Sarkar:11,Xu:2010}, and  formation of cracks~\cite{Allain:95,Dufresne:03,Gauthier:07,Inasawa:12,Xu:13}.

Confined directional drying has been used in many studies to better understand the drying of colloidal dispersions, from the mechanical instabilities mentioned above to transport phenomena~\cite{Sarkar:09,Goehring2017,Inoue2020,Hooiveld2023,Noirjean2017,Inasawa2016}. This is mainly due to the control of evaporation conditions in this one-dimensional drying geometry and the monitoring possibilities offered by this technique.
In addition to these studies,
confined directional drying is also a model configuration for studying the drying of porous media saturated with a colloidal dispersion in the capillary regime~\cite{Keita2013,Ghiringhelli2023}.
Despite all these studies, the evaporation kinetics, $J$ vs $t$,  and the growth of the porous solid, $x_c$ vs $t$, remain open questions to this day~\cite{Pingulkar2023}.  In the case of nanoparticles with typical radii $a \leq 20$~nm,  experiments have shown that the solid layer first grows linearly, $x_c \sim t$, and then follows a square root growth $x_c \sim t^{0.5}$ indicating  a decrease of the evaporation-induced flow  rate $J$~\cite{Dufresne:03,Sarkar:11,Inasawa:12,Lidon:14}. Most theoretical works
suggest that this slowdown is actually due to the minute recession of the evaporation interface within the porous solid, adding a high resistance to evaporation linked to the diffusion of water vapour in the dry porous layer, the so-called \textit{capillary-limited regime}~\cite{Dufresne:06,Lidon:14,Wallenstein:11,Pingulkar2023}. For very small colloids ($a < 10~$nm), the water pore pressure at the evaporation interface can decrease down to values that affect the partial pressure of water in the the gas phase, the so-called Kelvin effect~\cite{Thiery2017}. In this \textit{flow-limited regime}, there is no resistance to evaporation in the gas phase, and the slowdown is explained by the increasing resistance due to the flow through the growing porous solid~\cite{Pingulkar2023}. To date, no work has been done to  differentiate experimentally between these two regimes (except ref.~\cite{Vincent2016}, but for drying-induced permeation in  nanoporous media), nor has quantitative modelling been carried out, due to the lack of data describing the porous solid, in particular its hydraulic permeability $\kappa$ (m$^2$). Furthermore, all the theoretical studies are  based on one-dimensional modelling of transport phenomena, and do not take into account the possible role played by cracks and solid delamination on the solvent evaporation.

Our aim in the present work is to bring new insights on the confined directional drying of a dispersion of silica nanoparticles (mean radius $a = 11~$nm),  using  experiments in a 
poly(dimethylsiloxane) (PDMS) microfluidic chip. Our experiments 
first enable the directional drying of the colloidal dispersion in a highly confined geometry (cross-section
$h \times w = 30 \times 100~\mu$m$^2$, see Fig.~\ref{fig:SetupAirFlow}), 
while providing accurate measurements of both the evaporation kinetics $J$ vs $t$, and the growth rate of the solid, $x_c$ vs $t$. We also exploit the presence of hydrophilic filler in the PDMS formulation used, to estimate the water potential $\psi$ (Pa) along the porous solid as it grows. These original data, combined with a model taking into account the presence of an air film linked to delamination, enable us to estimate the  permeability  $\kappa$ of the porous solid and show that the slowdown observed in our configuration corresponds to the  \textit{capillary-limited regime}.

The present paper is organized as follows. In Section~\ref{sec:materialsMethods}, we present the microfluidic tools used and the experimental methods employed. In Section~\ref{sec:results}, we present the main results of our work, in particular concerning  the confined directional drying of a nanoparticle dispersion and its quantitative description.  In Section~\ref{sec:conclusions}, we conclude our work and suggest various perspectives.

\section{Materials and Methods \label{sec:materialsMethods}}

\subsection{A PDMS chip for confined directional drying}

Fig.~\ref{fig:SetupAirFlow} shows schematically the chip we designed to study the confined directional drying of  colloidal dispersions. 
We  used  standard soft lithography techniques to fabricate this  chip with Sylgard-184 PDMS (mass ratio curing agent/polymeric base $=1/10$). The channel network is composed of two inlets (I$_1$ and I$_2$) connected to a serpentine-like channel of length $L=43$~mm, and height$\times$width $h\times w=30\times 100~\mu$m$^2$. The latter leads to a higher channel (height $h_a = 70~\mu$m) in which air at an imposed relative humidity ($\mathrm{RH}$)  flows. The height step in the channel ($h \to h_a$) helps to trap an air/water meniscus when an aqueous colloidal dispersion is injected using either inlet I$_1$ or I$_2$~\cite{Choi2012,Choi2015a}.
Water evaporation from the air/water meniscus then drives a flow at a rate $J$ as long as the meniscus remains trapped, accumulating the colloids  up to the formation of a porous solid that finally invades the channel. 
 Many similar microfluidic experiments have been carried out, particularly in the context of passive pumping~\cite{Goedecke2002,Namasivayam2003} or for evaporative assembly of colloidal materials~\cite{Choi2012,Choi2015a}, see ref.~\cite{Bacchin2022} for a review.

In the absence of colloids, the evaporation rate under isothermal conditions can be estimated by\cite{Goedecke2002,Namasivayam2003,Bacchin2022}:
\begin{equation}
J = \frac{k c_\text{sat}}{\rho_w} (1-\mathrm{RH})\,,
\label{eq:evappurewater}
\end{equation}
with $c_\text{sat}$ (kg/m$^3$) the saturation vapor concentration in air, $\rho_w$ (kg/m$^3$) the mass density of liquid water, and  $k$ (m/s) a mass transport coefficient.
For diffusion-dominated transport for the water vapour  over the length scale $\lambda$ (see Fig.~\ref{fig:SetupAirFlow}), $k \simeq (h_a/h)\,D_w^\mathrm{air}/\lambda$ with $D_w^\mathrm{air}$ (m$^2$\,s$^{-1}$),  
the diffusion coefficient of the water vapour in air~\cite{Massman1998}. Eqn~(\ref{eq:evappurewater})  leads to evaporation rates ranging from $J \simeq 0.5$ to $2.3~\mu$m\,s$^{-1}$
for relative humidity ranging from $\mathrm{RH} = 0.8$ to $0$.
However, the PDMS matrix is not totally impermeable to water, as water molecules can solubilize in the elastomer, diffuse, and then evaporate into the ambient air, a mechanism known as pervaporation. Water pervaporation in our PDMS chip
inevitably leads to  a flow directed towards the fixed air/water meniscus~\cite{Randall:05,Verneuil:04}, and thus superimposed on the flow induced by evaporation. This pervaporation-induced flow can in turn also concentrate the colloids as in the many applications reviewed in Ref.~\cite{Bacchin2022}.
Dollet et al.\ calculated the flow due to pervaporation for a single linear channel of rectangular cross-section~\cite{Dollet2019}. This relation leads to an estimated pervaporation-induced flow $J_p \simeq 6~\mu$m\,s$^{-1}$ for the filled serpentine channel shown in Fig.~\ref{fig:SetupAirFlow}, $H= 200~\mu$m, and an external relative humidity $\mathrm{RH}=0.5$.
 This flow, significantly larger than the evaporation-induced flow predicted by eqn~(\ref{eq:evappurewater}), can be reduced by increasing $H$, but not completely~\cite{Bacchin2022}. Indeed, for the serpentine channel, the assumption of a single linear channel no longer holds for large $H$,
 but theoretical estimates in this case~\cite{Noblin:08} still lead to  $J_p \simeq 0.1$--$1~\mu$m\,s$^{-1}$ for thicknesses $H$ of a few mm. 
 Hence, in order to fully neglect pervaporation in our work, the whole chip is immersed in water before and during the experiments, as demonstrated in ref.~\cite{Randall:05}. With this simple technique, the pervaporation-induced flows  are fully  eliminated after a time scale of the order of $\tau_p \simeq H^2/D_w^p$, $D_w^p \simeq 6$--$8\times 10^{-10}$~m$^2$\,s$^{-1}$ being the diffusion coefficient of water in PDMS~\cite{Bacchin2022}. This led us to choose a thickness $H \simeq 200~\mu$m for our experiments,
 a compromise between the time required to stop pervaporation,  $\tau_p$ of the order of a few minutes in this case,   and the limits of microfabrication of  thin PDMS chips.

\subsection{Confined directional drying experiments}
We studied a charge-stabilized dispersion of silica nanoparticles in water commercialized under the name Ludox AS40
(monodisperse anionic
grade, Sigma Aldrich). The volume fraction of the commercial
dispersion estimated using   dry
extract measurements is
$\varphi_0\simeq 0.24$ and 
the mean radius of the nanoparticles is $a = 11$~nm\cite{Sobac2020}.
All experiments  were carried out at room temperature, $T \simeq 22^\circ$C. 

In a typical experiment, the microfluidic chip is first  immersed in a water bath for about $1$~h, and an air stream of controlled $\mathrm{RH}$  is imposed  in the air channel (typical rate $0.4$~mL\,min$^{-1}$, HumiSys LF $\mathrm{RH}$ generator, InstruQuest Inc.). Then, the  dispersion of silica nanoparticles is 
injected in the chip through inlet I$_1$ using a small amount of excess pressure (typically 50~mbar, MFCS, Fluigent). The dispersion invades the main channel and the geometric step traps   the air/dispersion meniscus ($x=0$).
Then, a fluorinated oil (Fluorinert FC40, 3M) is gently injected from inlet I$_1$ towards I$_2$, leading  to the formation of an oil/dispersion meniscus at $x_m = L$, the inlet of the serpentine channel. This is only possible because our device has two inlets, I$_1$ and I$_2$, connected to the same main channel.  
Finally, we used bright field microscopy (Olympus IX73) at 2X magnification  (spatial resolution  $3.25~\mu$m/pixel) using a sCMOS camera (Orca 4, Hamamatsu) to track both the drying process along the channel and estimate the evaporation-induced flow rate from the displacement of the oil/dispersion meniscus by :
\begin{equation}
J = -\frac{\text{d} x_m}{\text{d}t}\,. \label{eq:Jxm}
\end{equation}

\subsection{Raman microspectroscopy}

The confocal Raman measurements shown in Section~\ref{sec:cluster} were obtained
with a custom-made Raman microspectrometer setup coupled to an inverted microscope (Olympus IX71).
A laser beam (wavelength $532$~nm, output power $\simeq 30$ mW) is focused with a water-immersion 60X objective (NA of $1.2$) on the water clusters   observed in the PDMS film and discussed in Section~\ref{sec:cluster}. The scattered light is collected by the same objective, filtered, and directed
to the spectrometer (Andor Shamrock, grating $600$~lines\,mm$^{-1}$, input slit width~$100~\mu$m). A confocal pinhole 
($100~\mu$m) reduces the out-of-focus contributions. Typical acquisition times were of the order of $2$~s.

\section{Results \label{sec:results}}
\subsection{Confined directional drying, global views}
\begin{figure}[htbp]
    \centering
    \includegraphics{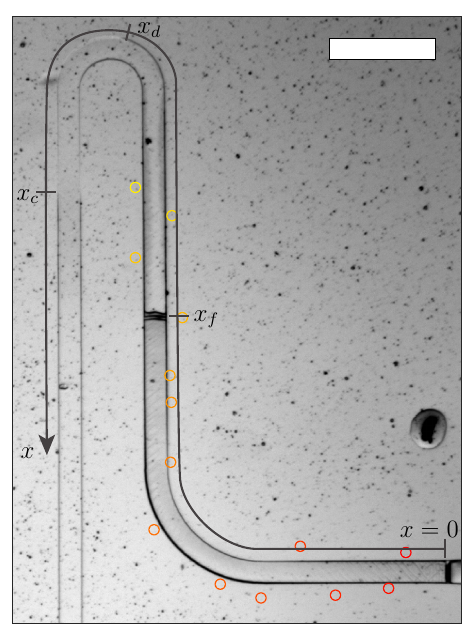}
    \caption{Snapshot at $t = 100~$min zooming on the first turn in the serpentine channel,
    see also movie M2.avi~\dag.
    $x_c$ indicates the compaction front, $x_d$ the delamination front, and $x_f$ the position of the single crack at that instant. The scale bar is equal to  $500~\mu$m. 
    The colored circles surround the positions of the water clusters
    in PDMS that are used to estimate the water potential, see Section~\ref{sec:cluster} and Fig.~\ref{fig:Impurity} (they have almost all disappeared at $t=100$~min). Their colors, from red to yellow, indicate their increasing  positions $x_i$ from the end of the channel, see the axis.}
   \label{fig:SnapShotAxes}
\end{figure}

Fig.~\ref{fig:SnapShotAxes} shows a snapshot at $t=100$~min of the confined directional drying of the dispersion of silica nanoparticles for an experiment performed with an air stream of relative humidity $\mathrm{RH}=0.2$,  see the movies M1.avi and M2.avi~\dag. For the sake of clarity, this snapshot zooms on the first turn in the serpentine channel, but the full field of view makes it possible to visualize the oil/dispersion meniscus  and thus measure the evaporation rate~$J$ using eqn~(\ref{eq:Jxm}).   

Our observations once again reveal the phenomenology commonly observed during the confined directional drying of such dispersions~\cite{Dufresne:03,Dufresne:06,Lidon:14}. At initial times, the accumulation of the nanoparticles  at $x=0$, leads 
to the formation of a porous solid
which then invades the main channel, as it is constantly fed  by an evaporation-induced flux of colloids.  A compaction front separates the dilute dispersion from this colloidal material. Its position $x_c$  is easily identified by the disappearance of the channel walls, due to     the matching of the optical indices of the PDMS and the dispersion at the compaction front. This transition from a liquid dispersion to a solid  is  accompanied by the onset of mechanical stresses~\cite{Wallenstein:11,Style2011}. In our case, these 
stresses deform the microfluidic channel, until the colloidal solid  detaches from it, as often reported for similar experiments~\cite{Xu:2010,Sarkar:11}.  The delamination front, denoted $x_d$, broadly follows the compaction front but intermittently, and is easily detected thanks to the strong optical contrast associated with the formation of an air film. In the experiment, see Fig.~\ref{fig:SnapShotAxes}, the solid initially detaches from the vertical walls of the channel, then later across its entire width. Subsequently, these mechanical stresses cause cracks to form in the colloidal solid, as also observed many times for similar experiments~\cite{Allain:95,Dufresne:03,Inasawa:12,Gauthier:07}, but with specific characteristics linked to the high confinement of the channel. The number of cracks is indeed relatively limited (3 in the experiment presented at $t= 500~$min), and they are transverse to the porous solid, literally cutting it  into  several distinct pieces. Even at 2X magnification, our observations  also revealed the presence of shear bands in the growing solid, as often described in similar experiments~\cite{Boulogne:14,Yang2015,Kiatkirakajorn:15}.

\begin{figure}[htbp]
    \centering    \includegraphics{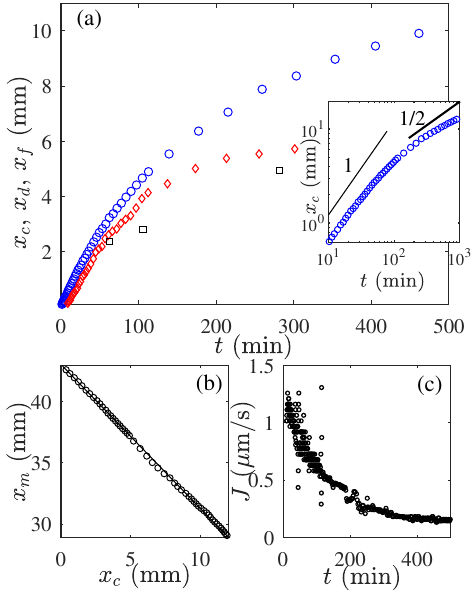}
    \caption{(a) Compaction front $x_c$ ({\color{blue} $\circ$}), delamination front $x_d$ ({\color{red} $\diamond$}) and  positions of cracks $x_f$ as they appear ($\square$) as a function of time $t$. Inset: $x_c$ vs $t$ in log-log scale. (b) Position of the meniscus~$x_m$ against the compaction front~$x_c$. The continuous line is a linear fit. (c) Evaporation rate $J$ estimated from the displacement of the meniscus using eqn~(\ref{eq:Jxm}). }
    \label{fig:GrowthEvapRate}
\end{figure}

Fig.~\ref{fig:GrowthEvapRate}a gathers the tracking of both the compaction and delamination fronts, $x_c$ and $x_d$ respectively, along with the positions  $x_f$ of the cracks when they appear. Initially, the compaction front grows linearly along the channel, $x_c \sim t$, and then slows down for $t \gtrsim 80$~min following at long time scale $x_c \sim t^\alpha$, with $\alpha \simeq 0.4$. This smooth growth does not seem to be affected by the  intermittent dynamics of the delamination front, nor by the sudden appearance of cracks. In addition, other experiments carried out under the same conditions led to very similar observations for $x_c$ vs $t$, but with noticeable differences for the  delamination front  and the number and positions of  the cracks, consistently with the stochastic nature of these phenomena. 

Fig.~\ref{fig:GrowthEvapRate}b shows the position of the oil/dispersion meniscus $x_m$, versus the compaction front $x_c$. The relationship between these two quantities is affine, and linked to the colloid conservation as discussed later.   Finally, Fig.~\ref{fig:GrowthEvapRate}c  shows the evaporation rate $J$ estimated from eqn~(\ref{eq:Jxm}) as a function of time. This curve shows that $J$ decreases continuously  from $J \simeq 1.3~\mu$m\,s$^{-1}$
to around $J \simeq 100~$nm\,s$^{-1}$ at long times.
All the observations described above, as well as the values of the dynamics $x_c$ vs $t$ and $J$ vs $t$ shown in Fig.~\ref{fig:GrowthEvapRate} are consistent with published data for similar experiments~\cite{Dufresne:03,Dufresne:06,Sarkar:11,Inasawa:12}, see in particular ref.~\cite{Lidon:14}. On the other hand, it is important to point out that our data suggest a slowdown at long times following $x_c \sim t^\alpha$, with an exponent $\alpha \simeq 0.4$ slightly lower than the $0.5$ value often reported in the references cited above. Although it is always tricky to estimate such exponent from fits over small time ranges, we believe that this over-slowdown is significant and possibly due  to the permeation of water from the  reservoir on top of the PDMS chip towards the channel,  as discussed in Section~\ref{sec:discussions}.  

\subsection{Water clusters in PDMS \label{sec:cluster}}

\begin{figure}[htbp]
    \centering    \includegraphics{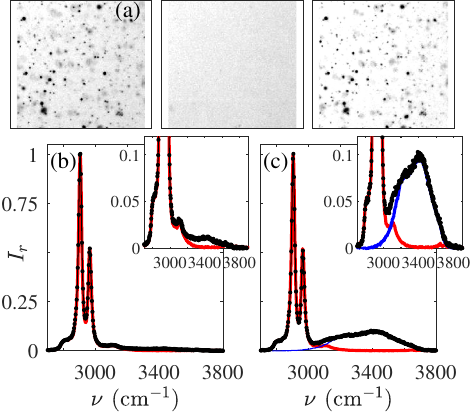}
    \caption{(a) Images ($325 \times 325$~$\mu$m$^2$) of a PDMS layer  $200~\mu$m thick, which, from left to right, is first immersed in a water bath, then removed and exposed to 
    ambient air, before being immersed again in the water bath.  
    (b) and (c): Raman spectra ($\cdot$) measured at the position of a water cluster in a PDMS film in (b) ambient air and (c) when immersed in water. Both spectra are normalized by the Raman peak  at $\nu \simeq 2900$~cm$^{-1}$ due the PDMS matrix. The red curves correspond to the Raman spectrum of PDMS measured in a cluster-free region, and blue curves to the spectrum of water. The insets show zooms  on the OH stretching vibrations in the spectral range $\nu = [3000$--$3800]$~cm$^{-1}$.}
    \label{fig:RamanDots}
\end{figure}

In addition to the observations described above, our experiments revealed a new phenomenon. When the chip is immersed in the water bath, a large number of small inclusions
appear in the PDMS, absorbing or scattering light, therefore suggesting impurities, see the black dots on Fig.~\ref{fig:SnapShotAxes}. 
Local observations at higher magnifications show that these inclusions are highly dispersed, with sizes ranging from  fractions of a micron to nearly 10~$\mu$m for the largest. When the chip is removed from the water bath, these inclusions disappear, and the PDMS becomes completely transparent again. As shown in Fig.~\ref{fig:RamanDots}a, the inclusions appear again exactly at the same positions as before when the chip is re-immersed in water, suggesting the presence of preferential sites within the PDMS matrix. 
Confocal Raman spectroscopy measurements on these inclusions revealed the OH stretching vibrations of the water molecules, suggesting that they are in fact microscopic water clusters,
see Fig.~\ref{fig:RamanDots}b and~\ref{fig:RamanDots}c.

Such observations are not new, as silicone elastomers are known to become milky when immersed in water, due to  the formation of water clusters that scatter light.
The origin of these clusters is actually linked to the presence in the  matrix of hydrophilic sites, either  impurities or filler materials, which can significantly absorb water molecules~\cite{Watson1996}. In the case of Sylgard-184, the poly(dimethylsiloxane) used in the present study, these sites  are probably silica filler with high specific surface area, added to the commercial blend 
by the manufacturer
to improve the mechanical properties of the  elastomer~\cite{Harley2012}.    
Harley et al.\ have measured  the role of these hydrophilic sites on the equilibrium sorption of water vapour in cross-linked  Sylgard-184~\cite{Harley2012}. For low water vapour activity ($\mathrm{RH} < 0.1$), these sites allow the immobilization of water molecules via Langmuir-type adsorption, whereas the sorption of water increases drastically at 
 higher water vapour activity, $\mathrm{RH}>0.7$--$0.8$.
 These authors
 attributed this effect to the clustering of the water molecules around the  Langmuir sites. 
These results led us to the  possibility that the water clusters we observe are also sensitive to the water activity $a_w$ in which the PDMS is immersed, and thus on the water potential $\psi = RT/V_m\log(a_w)$, with $V_m \simeq 1.805 \times 10^{-5}$~m$^3$mol$^{-1}$,  the liquid water molar volume at room temperature ($T = 22^\circ$C), and $R$ the universal gas constant.

\begin{figure}
    \centering
    \includegraphics{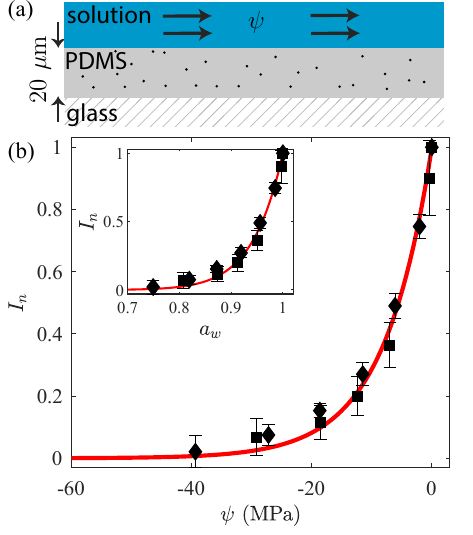}
    \caption{(a) Sketch of the calibration experiment of the water clusters in PDMS.
   A solution at a given water potential $\psi$  flows in a wide channel (height $\simeq 200~\mu$m, width $\simeq 1$ mm) over a thin layer of PDMS bonded on a glass slide ($\simeq 20~\mu$m). The drawing shows only the bottom PDMS layer.
   (b) Relative normalized intensity $I_n$ against $\psi$ for glycerol solution ($\blacklozenge$)  and NaCl solutions ($\blacksquare$). The red curve is a fit by eqn~(\ref{eq:calib}) with 
    $\psi_c \simeq -8~$MPa. Errorbars show the standard deviations based on data from eight different water clusters. 
    Inset: same data against the corresponding water activity $a_w = \exp[V_m \psi / (RT)]$.}
    \label{fig:CalibrationDots}
\end{figure}
Fig.~\ref{fig:CalibrationDots}a
shows the experiments we did for testing this assumption. A thin PDMS film is coated on a glass slide, and we added  a wide channel ($\simeq 1~$mm in width, $\simeq 200~\mu$m in height) on top
to inject aqueous solutions of known water potential,
binary mixtures
water/NaCl (at molalities up to $5$~mol/kg)~\cite{Fontana2007} and water/glycerol (at glycerol mass fractions up to $0.6$)~\cite{Ninni2000,Bouchaudy:18}. 
For $\psi \leq -40$~MPa ($a_w \leq 0.75$), we do not observe any water clusters at the magnification 10X, while the  clusters grow in size when the water potential increases  up to $\psi = 0$ ($a_w = 1$, pure water). For this thickness of PDMS layer, $\simeq 20~\mu$m, we did not observe any kinetic effect as the  appearance or disappearance of the clusters follows the imposed water potential  almost instantaneously.
To quantify more precisely these observations, we averaged the light intensity collected by the camera over a few microns around several clusters, then subtracted the intensity measured in a cluster-free region to eliminate 
 the fluctuations of the illumination.
This quantity is finally normalized between $0$ and $1$ and plotted in Fig.~\ref{fig:CalibrationDots}b against~$\psi$. 
Despite the errorbars, mainly due to the dispersity of the observed inclusions,
these measurements show  that the water clusters are indeed sensitive to the water potential, as the normalized intensity $I_n$ can roughly be fitted by:
\begin{eqnarray}
I_n = \exp(-\psi/\psi_c)\,, \label{eq:calib}
\end{eqnarray}
with $\psi_c \simeq -8~$MPa. 
This range of water potentials and of corresponding water activity $a_w$ is consistent with the observations of Harley et al.~\cite{Harley2012} that have reported a drastic increase of the water solubility in Sylgard-184 PDMS for 
vapour activity $\mathrm{RH} >0.7$--$0.8$. 
As noted by Harley et al.~\cite{Harley2012},  the exact mechanism for the  water clustering around the hydrophilic sites, and its dependence with the water potential remain unknown.  This mechanism could be similar to  the osmotic pumping mechanism that has been proposed  
 to explain the high water uptake of vulcanized rubbers, which is also due  to the  clustering of water molecules around hydrophilic impurities~\cite{Fedors1980,Thomas1987}. However, the origin of an osmotic force induced by the hydration of fillers in Sylgard-184 remains to be demonstrated, and the role of  the PDMS elastomer elasticity remains to be investigated.

\subsection{Local estimates of the water potential\label{sec:localest}}
\begin{figure}
    \centering
    \includegraphics{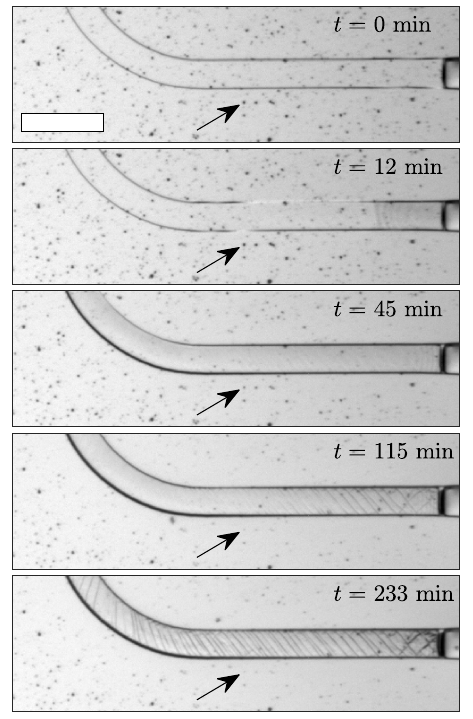}
    \caption{Snapshots at different times $t$ zooming in on the beginning  of the serpentine channel, see also  movie M3.avi~\dag.
    Arrows point to a water cluster that disappears at $t \simeq 100~$min.
    The scale bar is equal to $300~\mu$m.
    The change of the texture of the solid at its end for $t \gtrsim 100$~min reveals the shear bands initially present, and suggests the invasion of air/water nanomenisci, see Section~\ref{sec:applexp}.
   }
   \label{fig:ZoomGrowth}
\end{figure}

\begin{figure}
    \centering
    \includegraphics{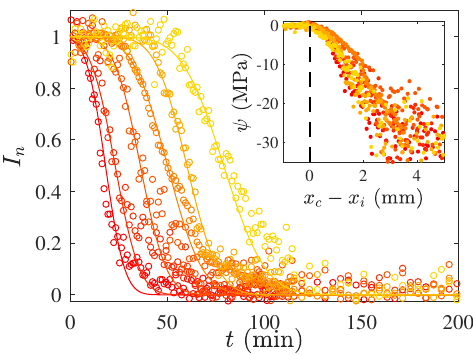}
    \caption{Relative normalized intensity $I_n$ against time $t$ for the water clusters shown in Fig.~\ref{fig:SnapShotAxes}. Only 6 curves are shown for the sake of clarity, and the color gradient indicates the increasing positions $x_i$ of the clusters along the channel (from $x_i \simeq 0.2$ to $3$~mm). The continuous lines are guides for the eye. 
    Inset: Corresponding  water potential $\psi$ using the calibration given by eqn~(\ref{eq:calib}) with $\psi_c = -8$~MPa, vs $x_c - x_i$ for the positions $x_i$ shown in the main graph.}
    \label{fig:Impurity}
\end{figure}

In the following, we exploit these clusters as  sensors of the  water potential in the PDMS matrix. Fig.~\ref{fig:ZoomGrowth} shows several snapshots zooming at the end of the evaporation zone, $x=0$, evidencing the growth of the colloidal solid at initial times. 
These images also reveal that some of the initially visible inclusions near the channel disappear over time, as the solid invades the channel. 
To better quantify this effect, we have selected 14 water clusters that disappear with time, located very close to the channel and   with positions  along the channel ranging from $x_i= 0.2$ to $3$~mm, see the colored circles in Fig.~\ref{fig:SnapShotAxes}. As for the calibration shown in Fig.~\ref{fig:CalibrationDots}, we have calculated the  normalized intensity $I_n$ measured by the camera   for each water cluster, see Fig.~\ref{fig:Impurity}. 
This plot clearly evidences a front of disappearance of these inclusions, parallel to the progression of the porous solid
in the channel. 

In order to make a rough estimate of the water potential at the locations of these different clusters, these curves are converted into water potential using eqn~(\ref{eq:calib})  and plotted against $x_c-x_i$, see the inset of Fig.~\ref{fig:Impurity}. 
These data unambiguously show  that the water potential in the vicinity of the microfluidic channel at a position $x_i$
 decreases over time as the colloidal solid  grows beyond this position in the channel (i.e., when $x_c - x_i > 0$). 
In the following, we will use these data to estimate the water potential $\psi$ at the different positions $x_i$ along the channel.

\subsection{Model}

\begin{figure}
    \centering
    \includegraphics{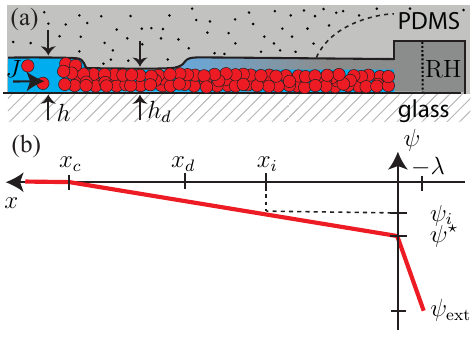}
    \caption{(a) Schematic view of the transverse section along the serpentine channel showing both the compaction and delamination fronts at a given time $t$ (not to the scale). The dotted line
    shows the cluster-free region in PDMS (black dots), where the  water potential $\psi \ll \psi_c= -8$~MPa.    
    (b) Corresponding profile  $\psi$ vs $x$  at time $t$.
    $x_i$ shows the position of a given cluster and its water potential $\psi_i$. $\psi_\text{ext} = RT/V_m \log(\mathrm{RH})$ is the water potential in the air stream of imposed relative humidity $\mathrm{RH}$, and $\psi^\star = \psi(x=0)$.}
    \label{fig:SketchModelColloid}
\end{figure}

We now present a simplified model of the directional drying in our experiments. 
Fig.~\ref{fig:SketchModelColloid}a shows a schematic view along the channel of the growth of the porous solid at a given time~$t$.
In this picture, we do not consider the concentrated layer of unconsolidated colloids upstream the compaction front ($x>x_c$) even if the later  can extend over large scales due to collective diffusion in such charge-stabilized dispersions~\cite{Goehring2017,Sobac2020}. The observed delamination implies that the solid has a smaller cross-section than the channel cross-section $(h \times w)$. In the following, we neglect the small shrinkage of the solid across its width (a few microns for $w = 100~\mu$m) and we assume that most of the shrinkage occurs across the height as also observed in ref.~\cite{Sarkar:11} and~\cite{Xu:2010}. This is mainly due to the relative high aspect ratio of the channel ($h \times w = 30 \times 100~\mu$m$^2$) and the  small thickness of the PDMS chip ($H = 200~\mu$m). 

In the configuration sketched in Fig.~\ref{fig:SketchModelColloid}, the colloid mass balance is:
\begin{equation}
(h_d\varphi_d - h\varphi_0) \frac{\mathrm{d} x_c}{\mathrm{d} t} \simeq h\varphi_0\,J\,, \label{eq:volumeconservation}
\end{equation}
$\varphi_d$ being the volume fraction of the porous solid, and $h_d$ its thickness, both assumed constant. This relation can be derived strictly from convection-diffusion models~\cite{Style2011}  (see also
 ref.~\cite{Pingulkar2023}  discussing its  validity for a non constant evaporation rate), and includes  the variation of height $h \to h_d$ at the compaction front. This relation is consistent with the  linear relationship observed between the compaction front $x_c$ and the position of the meniscus $x_m$, see Fig.~\ref{fig:GrowthEvapRate}b and eqn~(\ref{eq:Jxm}).
As shown by Lesaine et al.\ using precise dry extract measurements~\cite{Piroird2016,Lesaine2021}, the volume fraction of dried layers of similar Ludox dispersions is $\varphi_d \simeq 0.64$ for  evaporation rates larger than $\dot{E} \geq 10$~nm/s. Values close to $\varphi_d \simeq 0.64$  have also been measured using in situ small-angle X-ray scattering on similar nanosilica dispersions drying in similar configurations~\cite{Boulogne:14,Yang2018}.
In the following, we thus assume  the same value, and the linear fit shown in   Fig.~\ref{fig:GrowthEvapRate}b along with eqn~(\ref{eq:volumeconservation}) leads to  $h_d/h \simeq 0.82$. This value corresponds to a thickness of $\simeq 5~\mu$m for the air gap shown in Fig.~\ref{fig:SketchModelColloid}a, and is consistent with observations  at high magnifications of the dried solid across its height (data not shown). 

Estimating the rate of evaporation in the configuration shown in  Fig.~\ref{fig:SketchModelColloid}  a priori requires  a three-dimensional model   taking into account both the transport of the water vapour in the air gap downstream of the delamination front ($x<x_d$) and the  transport of liquid water through the porous solid ($x<x_c$). For this model, we  follow 
an approach similar to that used in
ref.~\cite{Vincent2016}  
 and~\cite{Pingulkar2023} based on the use of the water potential $\psi$ as it allows to describe water independently of its
state: liquid in the dispersion and in the porous
solid, to vapor in the gas phase. We  also assume local thermodynamic equilibrium and quasi-static conditions, i.e., continuity of the water potential. We also consider a porous solid fully saturated with water, and we do not initially  consider  the role played by cracks, see below for a discussion.

The  mass flux of liquid water $J^m_l$ (kg\,m$^{-2}$\,s$^{-1}$)  through the colloidal solid is driven by gradients of water potential and described by the  
Darcy's law:
\begin{eqnarray}
    J^m_l  = -\rho_w\frac{\kappa}{\eta_w}\nabla \psi\,,
    \label{eq:Darcy}
\end{eqnarray}
with $\kappa$ the permeability of the solid, and $\eta_w$ (Pa\,s) the  viscosity of water. 
In the gas phase, the mass flux of water $J^m_g$ (kg\,m$^{-2}$\,s$^{-1}$) is driven by  gradients of water vapour concentration $c_w$ (kg\,m$^{-3}$), and given by: 
\begin{eqnarray}
   J^m_g = - D_w^\text{air} \nabla c_w\,.  \label{eq:AirFluxMass}
\end{eqnarray}
This relationship can be rewritten using the definition of the water potential in the gas phase: 
\begin{eqnarray}
 \psi = \frac{R T}{V_m} \log\left(\frac{c_w}{c_\text{sat}}\right)\,,
\end{eqnarray}
leading to:
\begin{eqnarray}
    J^m_g = - D_w^\text{air} c_\text{sat} \frac{V_m}{RT}\exp\left( \frac{V_m \psi}{RT}\right) \nabla \psi\,. 
\label{eq:AirFluxMassPsi}
\end{eqnarray}
Continuity for the fluid flow (assuming constant density $\rho_w$) and quasi-static conditions in the gas phase imposes $\text{div}(J^m_l) = 0$ and $\text{div}(J^m_g) = 0$ respectively, and thus:
\begin{eqnarray}
&& \nabla \cdot \left[-\rho_w\frac{\kappa}{\eta_w}\nabla \psi \right] = 0\,, \label{eq:diffliq}\\
&& \nabla \cdot  \left[-D_w^\text{air} c_\text{sat} \frac{V_m}{RT} \exp\left( \frac{V_m \psi}{RT} \right)  \nabla \psi\right]  = 0\,. \label{eq:diffgaz}
\end{eqnarray}
Eqn~(\ref{eq:diffliq}) and (\ref{eq:diffgaz})  can be seen as two steady state  diffusion equations for the water potential $\psi$ with the effective diffusion coefficients,  $\rho_w \kappa/\eta_w$ for the liquid phase and
$D_w^\text{air} c_\text{sat}  \frac{V_m}{RT}\exp\left( \frac{V_m \psi}{RT} \right)$ for the vapour phase, the latter being a function of $\psi$.
Boundary conditions for these equations are  given by $\psi(x_c,t)=0$ (the liquid dispersion, neglecting  the osmotic contribution of the unconsolidated colloids upstream the compaction front), $\psi(-\lambda,t) = \psi_\mathrm{ext} = RT/V_m \log(\mathrm{RH})$ (imposed relative humidity in the air stream), and impermeability on the glass slide $\nabla \psi \cdot n = 0$. The boundary condition on the PDMS walls is less obvious as the PDMS matrix is itself permeable to water. We will assume for simplicity that the transport in the PDMS matrix is  negligible, and that the PDMS walls are almost impermeable (see Section~\ref{sec:discussions} for a discussion of this approximation).
Eqn~(\ref{eq:diffliq}) and (\ref{eq:diffgaz}) along with the boundary conditions given above can be solved numerically for a given geometric configuration,  such as the one shown schematically in Fig.~\ref{fig:SketchModelColloid}a. Such a numerical resolution should then lead to the mass fluxes given by eqn~(\ref{eq:Darcy}) and (\ref{eq:AirFluxMassPsi}), and thus to the global evaporation rate $J$. 
Nevertheless, we show below using simple theoretical arguments that the vapour transport through the air gap is negligible, and that the evaporation rate $J$ can be described by a simple one-dimensional equation. 

The ratio of the two effective diffusion coefficients  in eqn~(\ref{eq:diffliq}) and (\ref{eq:diffgaz}) is given by:
\begin{eqnarray}\epsilon =  \left. \left(\frac{D_w^\text{air}  c_\text{sat} V_m}{R T}\right) \exp\left( \frac{V_m \psi}{RT} \right) \middle/ \left(\frac{\rho_w\kappa}{\eta_w}\right)\,\right.\,. \label{eq:defepsilon}
\end{eqnarray}
This ratio compares the  mass  flux of water through the gas phase with that through the porous solid for a given water potential gradient. Assuming that the permeability $\kappa$ of the solid is correctly described by the Carman-Kozeny relation~\cite{Rogers2013}:
\begin{eqnarray}
\kappa_{CK} = \frac{(1-\varphi_d)^3}{45 \varphi_d^2}a^2 \,, \label{eq:Carman}
\end{eqnarray}
$\kappa_{CK}=3.1 \times 10^{-19}$~m$^2$, 
$\epsilon\ll 1$ even for high water potential ($\psi = 0$). 
Furthermore, the aspect ratio $h/x_d$ in the geometry shown in  Fig.~\ref{fig:SketchModelColloid}a is extremely small and the water potential is therefore approximately uniform  over the channel cross-section, i.e., $\psi \simeq \psi(x,t)$. 
Because $\epsilon \ll 1$,  the mass flux of liquid water through the porous solid
is much higher than the water mass flux in air, and the global mass balance imposes $(h w) \rho_w J \simeq (h_d w) J^m_l$. 
Eqn~(\ref{eq:Darcy}) then shows that the water potential
decreases linearly along the channel following the Darcy's law:
\begin{eqnarray}
\psi(x,t) \simeq \frac{h}{h_d} \frac{\eta_w J}{\kappa}  [x-x_c(t)]\,, \label{eq:psilinear}
\end{eqnarray}
as if the air gap due to delamination was not playing a role (negligible transport through the gas phase). This linear decrease is shown  schematically in Fig.~\ref{fig:SketchModelColloid}b. This assumption is also found in work on confined directional drying~\cite{Dufresne:03,Dufresne:06}, which implicitly assumes that the gas phase in the cracks and air films due to delamination is in equilibrium with the liquid phase inside the porous solid. It should be noted, however, that vapor in equilibrium in the air gap may have a humidity $\mathrm{RH}$ significantly smaller than $1$, due to the potentially low water potential values, see Section~\ref{sec:applexp}.

The simplified model presented above does not take into account the cracks observed in the porous solid, see Fig.~\ref{fig:SnapShotAxes}.  Yet these cracks literally cut the solid into several pieces, calling into question the description of a continuous liquid flow as described by Darcy's law, eqn~(\ref{eq:Darcy}). Nevertheless, it is quite conceivable that water is transported through the vapour phase between the different solid pieces. As the thickness $\xi$ of the cracks is very small compared to $x_c$, $\xi \simeq 20$--$40~\mu$m , eqn~(\ref{eq:diffliq}) and (\ref{eq:diffgaz}) can be used to show that the   resistance added by the cracks is negligible despite the fact that $\epsilon \ll 1$. This confirms the continuous description of eqn~(\ref{eq:psilinear}) and also probably explains why the appearance of cracks does not significantly affect the dynamics of both the compaction front $x_c$ and the evaporation rate $J$ in our experiments, see Fig.~\ref{fig:GrowthEvapRate}.

\subsection{Application to experiments \label{sec:applexp}}

\begin{figure}
    \centering
    \includegraphics{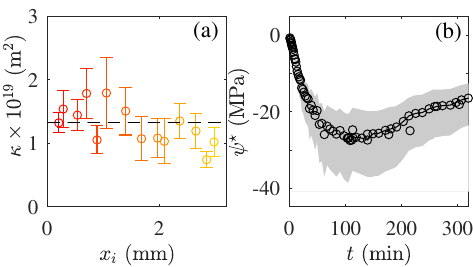}
    \caption{(a) Permeability values $\kappa$ estimated from
eqn~(\ref{eq:estimKappa}) as a function of the  position $x_i$ of the chosen cluster (see color code in Fig.~\ref{fig:SnapShotAxes}). The errorbars are calculated from the  standard deviation estimated for each cluster.  
     (b) Estimated water potential $\psi^\star$ at $x=0$ using eqn~(\ref{eq:psistar}) and $\kappa \simeq 1.3\times 10^{-19}$~m$^2$ ($\circ$). The shaded area corresponds to the same estimates but with a difference of $\pm 0.4\times 10^{-19}~$m$^2$ for $\kappa$.} 
    \label{fig:ModelExp}
\end{figure}

As demonstrated in Section~\ref{sec:localest}, the water clusters in PDMS allow  a rough estimate  of the water potential at their location $x_i$ along  the channel, see the inset of Fig.~\ref{fig:Impurity}. The clusters chosen in Fig.~\ref{fig:SnapShotAxes}  are located very close to the  side walls of the channel, so we can  assume they give good estimates of the water potential  in the channel  at  $x=x_i$.
 As described above, the water potential profile  decreases linearly along the channel  from $\psi(x_c,t) = 0$, see  eqn~(\ref{eq:psilinear}), so that we can estimate the permeability $\kappa$ from the measured values  $\psi(x_i,t)$ at the different clusters using: 
 \begin{eqnarray}
\kappa = \frac{h}{h_d} \frac{\eta_w J}{\psi(x_i,t)}  [x_i-x_c(t)]\,. \label{eq:estimKappa}
\end{eqnarray}
In order to eliminate the noise associated with very small values of water potential and $x_i-x_c$, we have only considered data from the inset in Fig.~\ref{fig:Impurity} ranging from  $\psi(x_i,t) =-25$ to $-5$~MPa to estimate $\kappa$ using eqn~(\ref{eq:estimKappa}) and the data on  evaporation rate $J$ and compaction front $x_c$ shown in Fig.~\ref{fig:GrowthEvapRate}. 
 This analysis leads to the $\kappa$ values shown in  Fig.~\ref{fig:ModelExp}a.
Despite the dispersion, no trend is observed in the permeability estimate, neither with the position $x_i$ of the different clusters, nor  with the time $t$ for a given cluster (not shown). Moreover,  we found the same trends and values for another experiment carried out under the same conditions, and it allows us  to reasonably confirm that $\kappa \simeq 1.3 \pm 0.4 \times 10^{-19}$~m$^2$. This  permeability value  is slightly smaller than the one estimated by the Carman-Kozeny relation, $ \kappa \simeq 0.4\kappa_{CK}$, possibly due to the polydispersity of the nanoparticles.
 It is worth mentioning that this lower value of permeability does not affect the hypothesis formulated above for the ratio of the two effective diffusion coefficients given in eqn~(\ref{eq:defepsilon}), as $\epsilon \ll 1$ even for this value of $\kappa$.  

Fig.~\ref{fig:ModelExp}b finally shows the extrapolated water potential at $x=0$:
\begin{eqnarray}
\psi^\star =  -\frac{h}{h_d} \frac{\eta_w J}{\kappa}  x_c\,,    \label{eq:psistar}
\end{eqnarray}
assuming  $\kappa=1.3 \times 10^{-19}$~m$^2$. 
Lidon et al.\ have also  reported similar measurements of the product $J x_c$ versus $t$ for similar dispersions~\cite{Lidon:14}, see  eqn~(\ref{eq:psistar}), but we go here a step further with our measurements of  $\kappa$ enabling us to estimate the water potential at $x=0$. 
$\psi^\star$ first decreases and  reaches a plateau for $t \gtrsim 80$~min  around $\psi^\star \simeq -27$~MPa, and then slightly grows for $t \geq 150$~min up to $\psi^\star \simeq -17~$MPa at $t \simeq 320~$min. The plateau  
 observed for $t \gtrsim 80$~min coincides with the transition between the linear growth rate for the compaction front,  $x_c \sim t$, and the slowdown observed on longer time scales, see Fig.~\ref{fig:GrowthEvapRate}a.

Also, this transition almost coincides with an apparent change of the texture of the solid material at its end, see the snapshots in Fig.~\ref{fig:ZoomGrowth}, and movie M3.avi~\dag. For $t \gtrsim 100$~min, the texture of the porous solid around $x = 0$ loses its homogeneity,  thus increasing 
the contrast of the   shear bands, which were barely visible at shorter times~\cite{Kiatkirakajorn:15,Yang2015}.
All these observations suggest that  the value $\psi^\star \simeq  -27$~MPa, much larger than the water potential imposed in the air stream $\psi_\text{ext} \simeq -220$~MPa,  actually corresponds to  the capillary pressure $\psi_\mathrm{cap}$ at which air/water nanomenisci invade the porous solid, leading to a higher optical contrast. It is traditionally assumed that this air invasion adds a resistance  to the water mass transfer, therefore slowing down the evaporation rate~\cite{Dufresne:06,Pingulkar2023}. Our quantitative estimate, $\psi_\mathrm{cap} \simeq -27$~MPa (with a range of confidence from $-37$ to $-20$~MPa because of the uncertainty on $\kappa$, see Fig.~\ref{fig:ModelExp}b) made possible thanks to our measurements of $\kappa$, is roughly consistent with standard estimates given by $\psi_\mathrm{cap}=  -\alpha \gamma/a$, with $\alpha$ a geometrical prefactor which, according to studies, leads to predictions ranging from
$\psi_\mathrm{cap} \simeq -35$~MPa as in ref.~\cite{Pingulkar2023} to
$\psi_\mathrm{cap} = -85~$MPa as in ref.~\cite{Dufresne:03}.

\subsection{Discussions \label{sec:discussions}}
As soon as air invades the porous solid, the description presented in Fig.~\ref{fig:SketchModelColloid} is no longer valid, as an additional resistance must be taken into account because of the transport  of the vapour by diffusion
across a dry layer of solid.  In particular, the extrapolated values of $\psi^\star$ using eqn~(\ref{eq:psistar}) are no longer valid either, and our model cannot therefore interpret the rise in $\psi^\star$ observed at long times in Fig.~\ref{fig:ModelExp}b ($t > 150~$min).
In this \textit{capillary-limited
regime}, i.e.,\ $\psi_\text{cap} > \psi_\text{ext}$,  the quantitative prediction of the evaporation rate remains an open question. 
All theoretical works~\cite{Wallenstein:11,Lidon:14,Pingulkar2023}
consider a one-dimensional description of directional drying.  These models suggest that desaturation takes place from the tip of the solid at $x=0$, and over a very small length due to the strong resistance to mass transfer by vapour diffusion through the porous dry layer.   
The change of the texture of the  solid at long times  shown  in Fig.~\ref{fig:ZoomGrowth} suggests that the invasion of the nanomenisci occurs over a long length scale within the channel, possibly leading to the progressive desaturation of the  porous solid across its height. These observations challenge the theoretical work cited above and call for new models and experimental measurements.

Finally, throughout our approach, we have  assumed that the channel walls are impermeable to water, but there may also be water transfer through the PDMS matrix. Considering typical values for water solubility and water diffusion coefficient in PDMS~\cite{Harley2012}, it is easy to show using a similar approach leading to  eqn~(\ref{eq:diffliq}) and (\ref{eq:diffgaz}) that the associated mass transfer resistance is very high, supporting our simplification. However, it is possible that mass transfer through the PDMS matrix plays a role at long times, when the driving force of evaporation along the channel becomes weaker and weaker ($J \to 0$). 
Indeed, the water bath imposes a water potential $\psi = 0$ at the top of the PDMS layer embedding the channel (thickness $H=200~\mu$m, see Fig.~\ref{fig:SetupAirFlow}). This implies that there is a minute flux of water from the bath towards the channel, as soon as $\psi(x,t) < 0$ in the channel. This flux could possibly explain the over-slowdown observed for the compaction front at long times ($x_c \sim t^{0.4}$), because it further reduces the 
 driving force of evaporation within the channel.
 All these considerations call for advanced, three-dimensional modeling of mass transfer in this type of experiment to confirm these assumptions.

\section{Conclusions and outlooks
\label{sec:conclusions}}
The model described in the previous sections, correlated with original measurements of the water potential, has enabled us to estimate both the Darcy permeability $\kappa$ and capillary pressure $\psi_\text{cap}$ of the porous solid made of silica nanoparticles. 
We are not aware of any measurements of these two quantities  by other independent techniques for comparison, but
the measured values are consistent with the theoretical predictions. 
Our experimental measurements open up a number of possibilities, 
in particular for the quantitative description of drying-induced stresses, as the value of $\kappa$  plays a fundamental role in poro-elastic modeling~\cite{Style2011,Chekchaki2013,Bouchaudy2019,Hennessy2022}.
In our experiments for instance, the knowledge of $\kappa$ enables us to estimate the water potential $\psi$, and therefore the pore pressure within the solid, a quantity related to the tensile component of the stresses during drying~\cite{Style2011,Bouchaudy2019,Hennessy2022}. The data shown in the inset of Fig.~\ref{fig:ModelExp} correlated with the observations gathered in Fig.~\ref{fig:GrowthEvapRate} can then be used to link the value of the pore pressure to the various observed instabilities: shear bands,  delamination, and cracks.

The quantitative description of confined directional drying presented in this work was made possible by original, but approximate, measurements of the water potential using (uncontrolled) hydrophilic sites within the PDMS matrix. Such observations have rarely been reported in the literature (in particular, the data in Fig.~\ref{fig:RamanDots} and~\ref{fig:CalibrationDots}) and open up numerous perspectives  that go well beyond the study of this work. We believe it is essential to better understand these phenomena, and even to use calibrated inclusions for more precise and controlled measurements of the water potential, because the significantly large errorbars of the calibration curve shown in  Fig.~\ref{fig:CalibrationDots} are likely due to the polydispersity of the  water clusters evidenced in our work. This approach shares  strong similarities with the methods developed by Jain et al.\ based on   fluorescent nanogels  in order to probe in situ  water potential in plant leaves with high precision~\cite{Jain2021}.  
It seems to us that these innovative approaches could also open up many new prospects for understanding drying of colloidal dispersions.

\section*{Conflicts of interest}
There are no conflicts to declare.

\section*{Acknowledgements}
We acknowledge Solvay, CNRS and the ANR program grant no. ANR-18-CE06-0021 for financial support.

\providecommand*{\mcitethebibliography}{\thebibliography}
\csname @ifundefined\endcsname{endmcitethebibliography}
{\let\endmcitethebibliography\endthebibliography}{}

\end{document}